 \documentclass[sigconf]{acmart}
\AtBeginDocument{%
  }

\copyrightyear{2026}
\acmYear{2026}
\setcopyright{cc}
\setcctype{by}
\acmConference[CI '26]{Proceedings of the ACM Collective Intelligence Conference}{September 27--30, 2026}{Alexandria, VA, USA}
\acmBooktitle{Proceedings of the ACM Collective Intelligence Conference (CI '26), September 27--30, 2026, Alexandria, VA, USA}
\acmDOI{10.1145/3834581.3838623}
\acmISBN{979-8-4007-2895-2/2026/09}

\begin{document}

\title[SleuthTalk]{SleuthTalk: Supporting Historical Photo Identification with Private Workspaces for Collective Sensemaking and Deliberation}



\author{Liling Yuan}
\orcid{0009-0003-1940-0629}
\affiliation{%
  \institution{Microsoft}
  \city{Redmond}
  \state{WA}
  \country{USA}
}
\email{lilingyuan@microsoft.com}
\authornote{These authors conducted substantial work on this project while at Virginia Tech.}

\author{Vikram Mohanty}
\orcid{0000-0001-6296-3134}
\affiliation{%
  \institution{\mbox{Human-Computer Interaction Institute}\\Carnegie Mellon University}
  \city{Pittsburgh}
  \state{PA}
  \country{USA}
}
\email{vikrammohanty@acm.org}
\authornotemark[1]

\author{Kurt Luther}
\orcid{0000-0003-1809-626}
\affiliation{%
  \institution{Department of Computer Science \& \mbox{Center for HCI}\\ Virginia Tech}
  \city{Alexandria}
  \state{VA}
  \country{USA}
}
\email{kluther@vt.edu}
\begin{abstract}

Identifying individuals in historical photographs is a critical task across fields such as history, journalism, genealogy, and archival research. While AI-based facial recognition can efficiently generate candidate matches, it often produces ambiguous results that require deeper analysis and contextual interpretation. Existing platforms lack robust support for collaborative deliberation, especially in uncertain or high-stakes cases. We present SleuthTalk, a private collaborative workspace integrated into Civil War Photo Sleuth, designed to scaffold structured comparison, discussion, and group decision-making. SleuthTalk enables users to curate custom shortlists, annotate facial features, and build consensus through structured feedback. In a mixed-methods evaluation with experienced historical photo researchers, SleuthTalk enhanced self-reported confidence, surfaced diverse perspectives, and supported transparent, reflective identifications.

\end{abstract}





\maketitle

\section{Introduction and Background}

Original historical research often resembles a \emph{socio-technical detective process}~\cite{collingwood1993idea,tosh2015pursuit,ginzburg2013clues}, requiring individuals to piece together fragmented evidence from distributed, often informal or incomplete sources~\cite{trouillot2015silencing}. Whether reconstructing a family tree on Ancestry.com or identifying individuals in historical photographs, these efforts demand domain expertise, contextual reasoning, and careful triangulation across records, images, and oral histories~\cite{luther2018non,luther2017merrill,10.2307/26925454}. In recent years, AI-based tools, such as computer vision algorithms and automated genealogical "hints," have significantly accelerated this work, helping researchers navigate vast archives and uncover previously lost connections~\cite{lim2023backtrace,mohanty2019photo,willever2014family}. However, the use of AI for original historical research remains far from perfect: misidentifications, ambiguous matches, and inaccurate family trees are common, especially when data is sparse or contexts are hard to verify~\cite{willever2014family,mohanty2020photo,mohanty2023photo}.

One illustrative domain is historical photo research, where computer vision techniques, including facial recognition and backdrop analysis, have enabled substantial progress in connecting unnamed portraits to known individuals, photographers, or locations~\cite{lim2023backtrace,mohanty2019photo}. Platforms like Newspaper Navigator and Civil War Photo Sleuth (CWPS) use content-based image retrieval (CBIR) to search large photographic databases and present users with shortlists of visually similar images~\cite{lee2020newspaper,mohanty2019photo}. These shortlists can be powerful starting points, but they are rarely conclusive~\cite{mohanty2019second}. In many cases, users must compare candidates that share strong visual resemblance and weigh subtle cues such as facial structure, military uniforms, biographical details, props, or backdrops. This interpretive step is often complicated by visual ambiguity, incomplete metadata, or lack of domain knowledge, which can lead to incorrect matches or unresolved identifications~\cite{harris2019civil}. In such moments, the algorithm’s role ends, and the responsibility shifts to the user to decide in a context where decisions are often made visible to others.

On platforms like CWPS, the final act of identification typically unfolds in public: users publish their proposed matches alongside supporting evidence, opening their reasoning to community scrutiny. While this transparency can foster collaboration~\cite{mohanty2023photo}, it also introduces social pressures that shape how and when people choose to share. Drawing on prior work in sociology and communication, we speculate that these pressures may lead to divergent behaviors. Some users, eager to contribute or appear knowledgeable, may offer confident identifications even in the face of lingering ambiguity. Others, wary of reputational risks or being publicly wrong, may choose not to publish a proposed match at all~\cite{bullingham2013presentation,marwick2011tweet}. Public self-presentation and impression management have all been shown to influence how individuals engage in high-visibility settings, particularly when expertise is at stake. Identifications on CWPS carry significant weight: incorrect conclusions have circulated in public history settings~\cite{luther2020real,luther2019gold}, contributed to false genealogies~\cite{willever2014family}, and influenced the perceived financial value of artifacts~\cite{handler2007retouching}. High-profile misidentifications beyond the history domain, such as the wrongful identification of Sunil Tripathi during the Boston Marathon bombing~\cite{starbird2014rumors}, or that of a retired firefighter falsely linked to the January 6 Capitol riot~\cite{harwell2021sleuths}, further illustrate how public visibility can amplify the consequences of error. These dynamics point to \textbf{the need for design approaches that support exploration, reflection, and revision—without requiring users to perform certainty in public.}

These dynamics are not only social, but structural. Popular platforms for collaborative historical research, such as Facebook groups, Reddit threads, and online forums~\cite{ruane_facebook_2014,gilbert2020run}, often lack the affordances necessary for sustained, reflective investigation. These environments are poorly suited to working through ambiguity because posts are buried quickly~\cite{hodas2012visibility,Wilson2024}, discussions can get long and fragmented~\cite{zhang2017wikum}, and contributions may be judged without context~\cite{marwick2011tweet}. In contrast, private, small-group settings offer an alternative: spaces where trusted collaborators, such as family, friends, or domain experts, can examine evidence together, consider competing hypotheses, and revise conclusions without fear of public missteps. Theoretical work on \emph{transactive memory systems} shows that such groups, when composed of individuals who understand and rely on each other’s complementary expertise, are more likely to reason effectively and reach confident outcomes~\cite{wegner1987transactive}. Similarly, structured deliberation has been shown to promote deeper engagement and better decisions, particularly when participants feel safe to express uncertainty or disagreement~\cite{kriplean_integrating_2014,kriplean_is_2012,lee2020solutionchat,mohanty2023photo}. Together, these perspectives point toward the value of shifting from \textbf{large-scale, open-ended crowdsourcing to smaller-scale, trust-based collaborative sensemaking.}

Our research in this paper builds on Civil War Photo Sleuth (CWPS)\footnote{\url{http://www.civilwarphotosleuth.com}}, a free online platform that uses AI-powered facial recognition and crowdsourcing to help users identify unknown Civil War-era portraits. CWPS enables users to upload a photo, search a database of over 60,000 images for visually similar matches, and evaluate potential candidates based on facial resemblance, uniform details, and historical records~\cite{mohanty2019photo}. While CWPS supports individual inquiry and public sharing, it offers limited affordances for trusted, collaborative deliberation. To address these challenges in a real-world setting, we extended CWPS with a new feature: \textit{SleuthTalk}, a private collaborative workspace integrated into CWPS. SleuthTalk allows users to curate custom shortlists of candidate matches, invite trusted collaborators into a shared deliberation space, and engage in structured sensemaking through feature-based comparisons, threaded discussions, and polling. Within each project, users can annotate facial features to support detailed visual comparisons, contribute external sources and contextual evidence, and use structured polls to collect group input and resolve disagreements. 

Our mixed-methods evaluation with experienced historical photo researchers found that SleuthTalk improved participants’ self-reported confidence in their identifications, supported balanced decision-making, and encouraged reflection on alternative hypotheses, compared to standard social media sites. Participants highlighted the value of seeing others’ reasoning processes, surfacing overlooked visual or contextual details, and being able to revise their own conclusions without judgment. By providing a space that is both private and persistent, SleuthTalk enables users to collaboratively weigh candidates without fear of premature exposure, share knowledge across domains, and reach ID decisions with greater confidence and care. Based on these findings, we discuss design opportunities for supporting trust-based, small-group deliberation in historical research and beyond. Our contributions include:

\begin{itemize}
    \item A collaborative deliberation workflow for addressing ambiguous decision-making tasks in historical person identification.
    \item A private, structured workspace, \emph{SleuthTalk}, integrated into the CWPS platform to support shortlist curation, feature-based comparison, and group discussion.
    \item A mixed-methods evaluation showing how SleuthTalk supports enhanced confidence, reflection, and collective sensemaking.
\end{itemize}

\section{System Design}

\textit{SleuthTalk} is designed to support deliberative collaboration around uncertain historical person identifications. We added SleuthTalk as a novel extension built upon the existing CWPS web platform. While CWPS uses face recognition and structured tagging to retrieve visually similar results, SleuthTalk extends this workflow by enabling users to curate intelligent shortlists, collaborate privately, and provide structured feedback during the identification process.

SleuthTalk consists of three core components: an intelligent interface for shortlisting potential candidates (Section~\ref{se:Shortlisting Potential Candidates}); a private collaborative workspace to invite and engage trusted collaborators (Section~\ref{se:Private Collaborative Workspace}); and structured feedback mechanisms for fine-grained visual comparison and group decision-making (Section~\ref{se:Structured Feedback}).

\subsection{Intelligent Shortlists: Curating and Managing Candidate Shortlists} \label{se:Shortlisting Potential Candidates}

After uploading an unknown Civil War photo, CWPS retrieves a list of facially similar candidates, ordered by similarity scores from Microsoft’s Cognitive Services Face API and filtered based on user-specified metadata~\cite{azure}. Users examine these candidates to determine if the facial resemblance and associated biographical data (e.g., service records) align with visual clues from the query photo (e.g., uniform insignia). However, prior work shows that correct matches are often ranked beyond the top-5 or even top-50 results~\cite{mohanty2020photo}. 

To address this, SleuthTalk allows users to add any promising candidates from the search results to a custom shortlist for deeper analysis. Each search result includes an “Add to Shortlist” button. When clicked, the candidate is added to a floating container on the right side of the screen, and the background of their result card is visually updated. Once the shortlist is finalized, users can create a new project by clicking the “Create Project” button.

\begin{figure*}[h]
    \includegraphics[width=\textwidth]{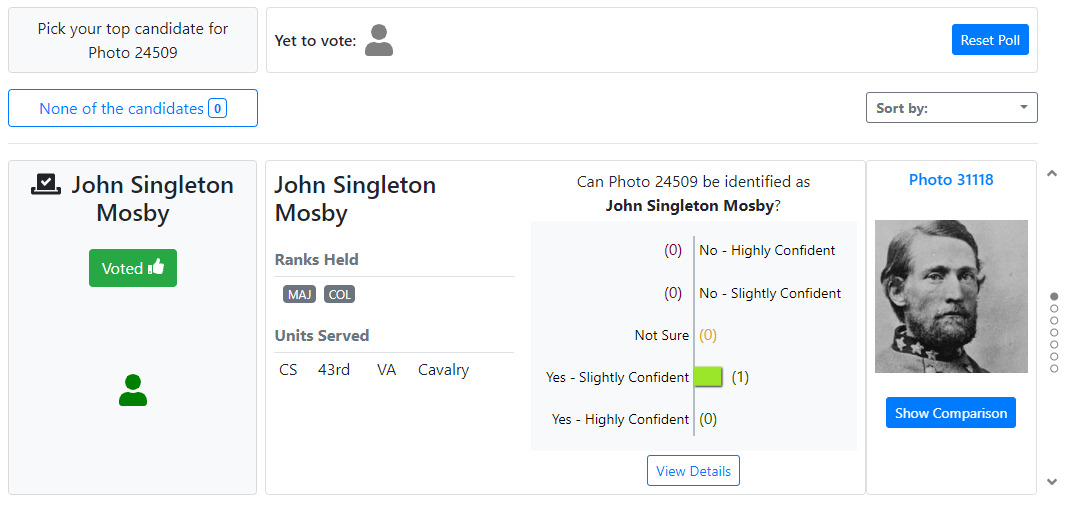}
    \caption{SleuthTalk candidate shortlist view. The interface displays (from right to left) a photo carousel with all known images of the candidate, structured identification vote options, a biographical summary including ranks and units served, and a visual summary of poll results. Users can click “Show Comparison” to review feature-level visual annotations and reasoning submitted by collaborators. (Public domain photo courtesy of the Library of Congress)}
    \Description{
    A screenshot of SleuthTalk’s candidate comparison interface. On the right is a photo carousel featuring a Civil War portrait labeled “Photo 31118” with a “Show Comparison” button beneath. In the center, a vote summary asks whether Photo 24509 can be identified as John Singleton Mosby, showing one slightly confident “yes” vote. The left panel highlights the candidate’s name, ranks (Major, Colonel), and military unit (43rd Virginia Cavalry, Confederate States), along with a green “Voted” icon. Interface controls for poll participation and sorting options are visible above the candidate panel.
    }
    \label{fig: poll_vote_section}
\end{figure*}

After project creation, all shortlisted candidates appear in the “Shortlisted Candidates” section of the project page. Each candidate card displays a photo thumbnail, biographical profile, and, if available, a vertical carousel of multiple images (see Figure~\ref{fig: poll_vote_section}). Bubble icons indicate the number of photos available. To mitigate position bias, the candidates are randomized by default, with optional re-sorting by time added or poll vote count.

After a project is created, any team member can add more candidates using the “Add Candidate” button and the search interface. Users may also archive (or unarchive) candidates from the shortlist. Archived candidates retain all previous votes and annotations and are accessible in the “Archived” section.

Each user may have one project per photo, enabling multiple shortlists and collaborative explorations of the same image across different social networks.

\subsection{Private Collaboration: Inviting Trusted Collaborators into a Shared Workspace} \label{se:Private Collaborative Workspace}

\subsubsection{Controlled Access}

Prior work found that CWPS users often seek second opinions from trusted individuals, especially family, friends, or domain experts, and express a desire to know the expertise behind identification claims~\cite{mohanty2019second}. SleuthTalk addresses this by offering private collaborative workspaces. The project creator can invite up to nine collaborators by username, name, or email, using an auto-complete interface. Non-CWPS users can also be invited via email. 

Previous work found that aggregated opinions from six anonymous crowd workers provided benefits~\cite{mohanty2019second}, although how many collaborators our participants would want remained an open question. We set the maximum at nine to exceed that baseline while balancing the diverse perspectives and expertise afforded by larger groups against the trust, privacy, and efficiency afforded by smaller ones.

Invited users can view basic project information but must accept the invitation to fully participate. Only project members can view or interact with shortlist votes, feature comparisons, or discussions. When the project creator finalizes a candidate, the identification becomes publicly visible on CWPS, but the discussion and deliberation remain private. This approach prevents premature misidentifications while allowing validated matches to benefit the broader community.

\subsubsection{Discussion Support} \label{se:Discussion Support}

In addition to structured voting, SleuthTalk includes a threaded discussion section for more open-ended conversation. Users can share contextual sources, ask questions, and explain disagreements. Specific candidates can be referenced using hashtags (e.g., “\#13249”), which display that candidate’s comparison information in a modal dialog (see Figure~\ref{fig: facial_feature_selection}).

\subsection{Structured Feedback: Comparing Features and Reaching Consensus} \label{se:Structured Feedback}

\subsubsection{Facial Feature Comparison} \label{se:Facial similarity comparison}

\begin{figure*}[h]
    \centerline{\includegraphics[width=\textwidth]{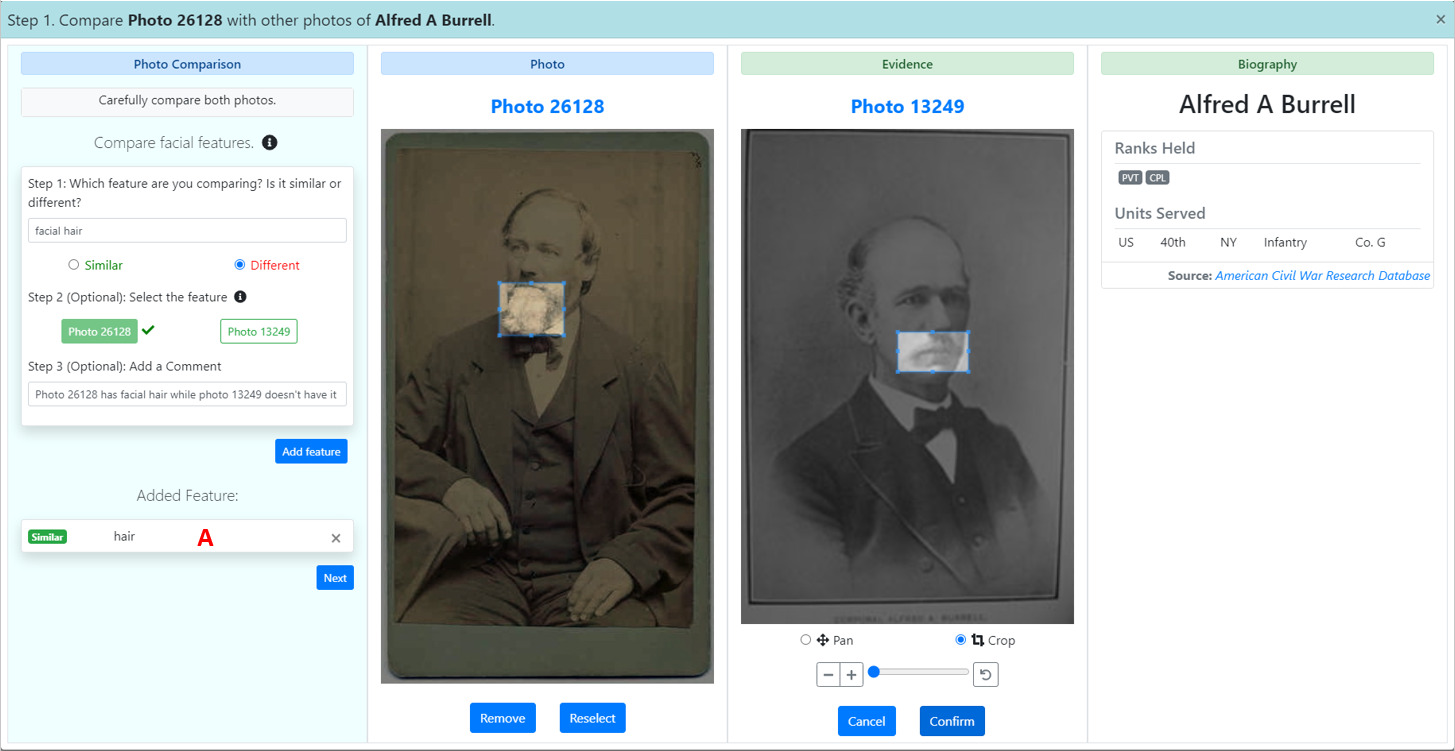}}
    \caption{SleuthTalk’s facial feature comparison workflow. Users begin by labeling a specific facial feature (e.g., hair, jawline, eyes), then crop the corresponding region from one or both photos using an interactive tool. They may optionally add a comment to contextualize the comparison. These steps support fine-grained visual analysis and can be repeated across multiple features to aid identification deliberation. (Public domain photos courtesy of the Robert Gray Collection [left] and American Civil War Research Database [right])}
    \Description{A screenshot of SleuthTalk’s photo comparison interface for identifying Alfred A. Burrell. The interface is divided into four columns: a control panel for selecting and labeling features, two side-by-side historical portraits (Photo 26128 and Photo 13249) with cropped regions overlaid on the lower face, and a biographical profile showing ranks and military unit. The left panel shows a labeled feature (“hair”) marked as different, with an added comment noting that one photo has facial hair while the other does not.}
    \label{fig: facial_feature_comparison}
\end{figure*}

CWPS originally included a two-step process: comparing facial similarity, followed by voting on identity confidence. Building on prior findings that facial feature comparisons improve user confidence~\cite{mohanty2019second}, SleuthTalk adds a third step for fine-grained comparison. Users identify, crop, and label individual facial features (e.g., eyes, nose, jawline) from both photos using a guided interface (see Figure~\ref{fig: facial_feature_comparison}).

\begin{figure*}[h]
    \centerline{\includegraphics[width=\textwidth]{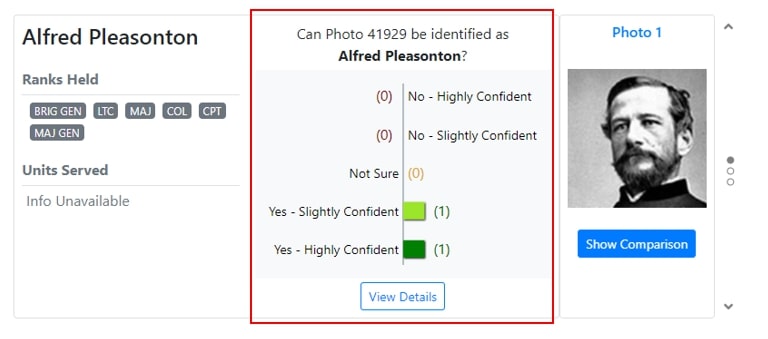}}
    \caption{Identification voting interface within a candidate card. SleuthTalk displays structured confidence levels—ranging from “No – Highly Confident” to “Yes – Highly Confident”—for determining whether the candidate matches the unknown photo. Users cast individual votes, which are aggregated into a vertical bar graph. Additional biographical details and access to facial comparisons are available via the adjacent profile and “Show Comparison” button. (Public domain photo courtesy of the Tony Mollo Collection)}
    \Description{Screenshot of a SleuthTalk candidate card for Alfred Pleasonton. The center panel shows a vote distribution asking if Photo 41929 matches this individual, with two votes cast—one “slightly confident” and one “highly confident.” The right panel displays a portrait labeled “Photo 1” with a “Show Comparison” button. The left panel includes the candidate’s name and a list of military ranks held; unit information is marked as unavailable.}
    \label{fig: Annotated_candidate_card}
\end{figure*}

\begin{figure*}[h]
    \centerline{\includegraphics[width=\textwidth]{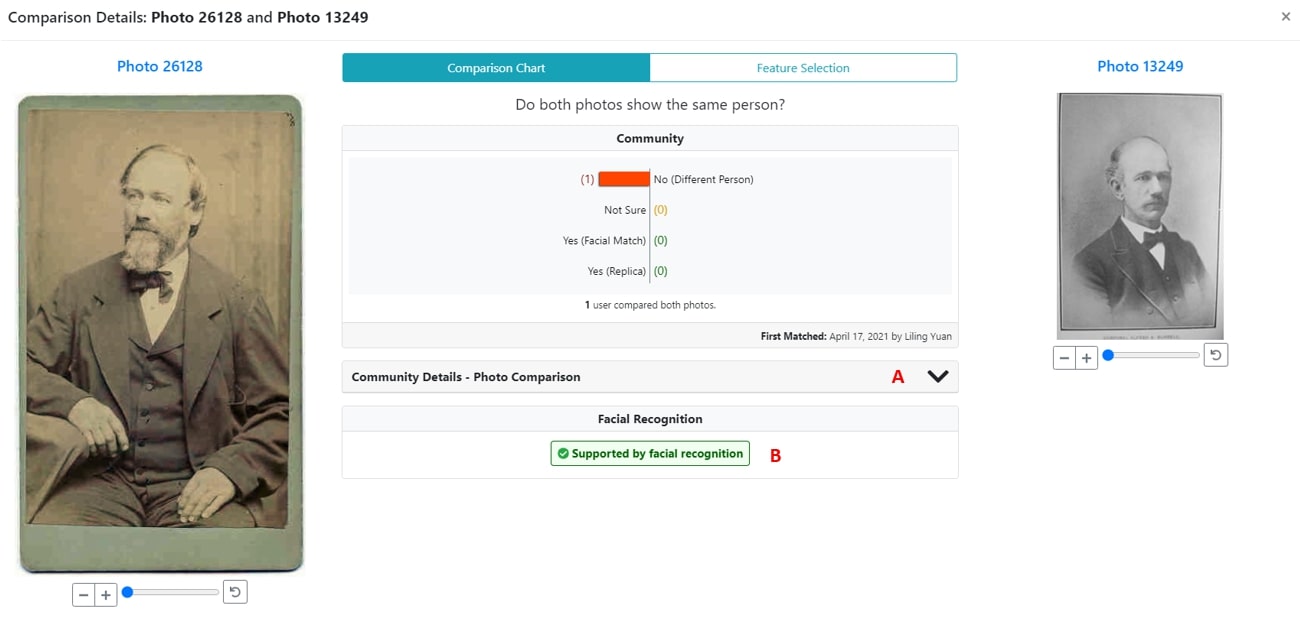}}
    \caption{Facial comparison summary between two photos. Section A aggregates community judgments about whether the two portraits depict the same person, with votes spanning from “No” to “Yes (Replica).” Section B indicates whether the match is supported by algorithmic facial recognition. This dual-layered feedback combines human interpretation with machine analysis to support informed decision-making. (Public domain photos courtesy of the Robert Gray Collection [left] and American Civil War Research Database [right])}
    \Description{Screenshot of SleuthTalk’s facial comparison chart between Photo 26128 and Photo 13249. On the left and right are historical portraits of two different individuals. The center panel presents voting results: one vote indicates the photos show different people. Below, section A is labeled “Community Details – Photo Comparison” with an expandable panel, and section B displays a green badge reading “Supported by facial recognition.” The match was first recorded in April 2021.}
    \label{fig: facial_comparison_chart}
\end{figure*}

An auto-complete dropdown ensures feature names are consistent across users~\cite{kittur2014standing}. After voting, a histogram visualizes user confidence in each candidate (see Figure~\ref{fig: Annotated_candidate_card}), and a comparison chart shows both human and algorithmic similarity assessments (see Figure~\ref{fig: facial_comparison_chart}).

\begin{figure*}[h]
    \centerline{\includegraphics[width=\textwidth]{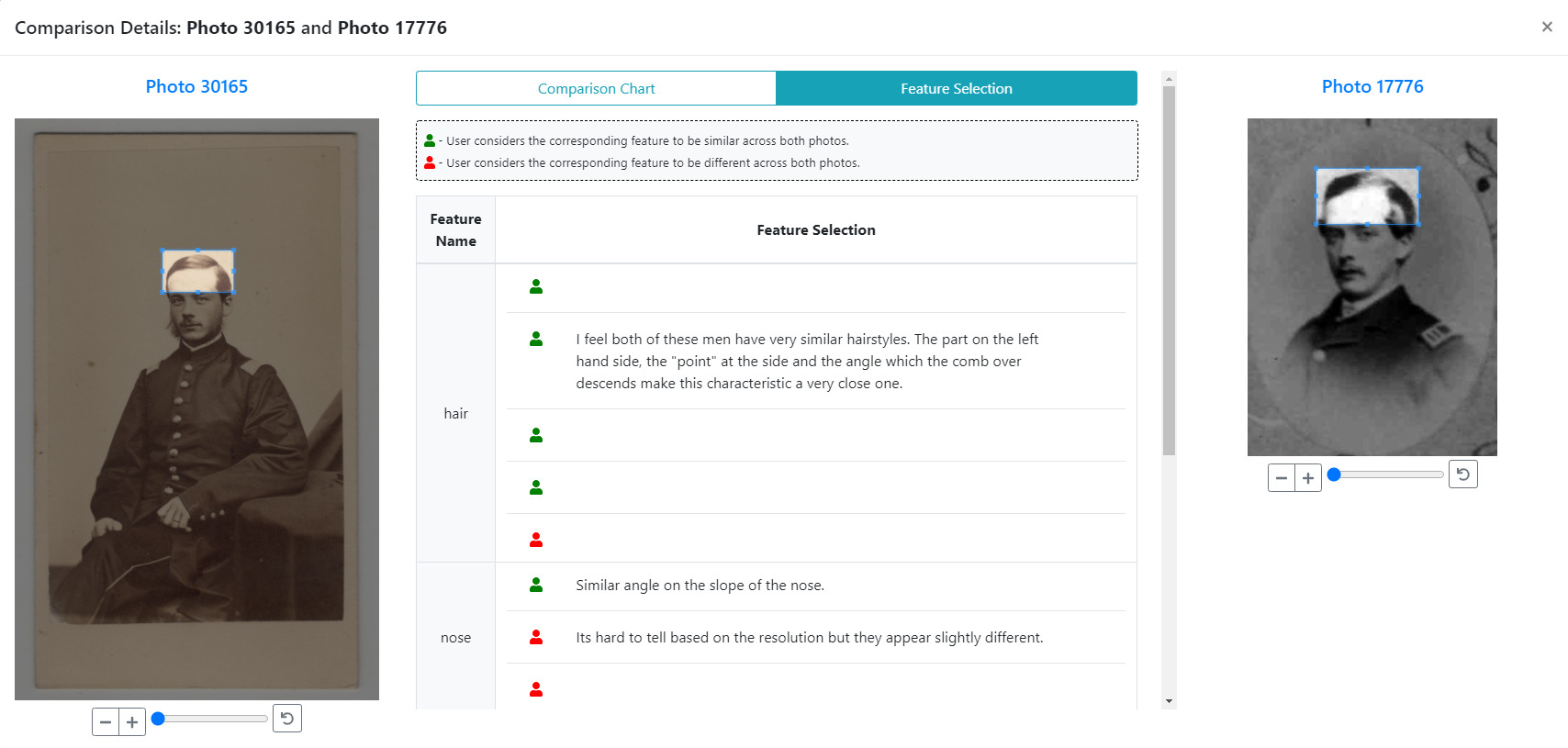}}
    \caption{Aggregated feature-level comparison results. Users crop and label specific facial regions in each image—such as hair or nose—and provide explanatory comments. The table displays a consolidated view of these annotations, showing which features were judged similar or different across the photos. Hovering over the cropped regions reveals precise areas of focus, supporting detailed, side-by-side visual analysis. (Public domain photos courtesy of the Isaac D. Friese Collection [left] and MOLLUS-Mass Collection [right])}
    \Description{Screenshot of SleuthTalk’s facial feature comparison interface for Photo 30165 and Photo 17776. Each portrait displays a rectangular crop over the top of the head. The central panel shows a feature selection table with two sections: “hair” and “nose.” Each row includes an icon indicating agreement or disagreement on similarity, and user-submitted comments. One note mentions comb direction and hairstyle similarity; another discusses nose slope differences. The tool offers structured insight into how users compare specific features between photos.}
    \label{fig: facial_feature_selection}
\end{figure*}

SleuthTalk also aggregates the feature-level comparisons of all project members, showing selected features, similarity annotations, crop regions, and optional comments (see Figure~\ref{fig: facial_feature_selection}).

\subsubsection{Identification Poll and Final Decision} \label{se:Identification Poll and Final Decision}

Once all candidates have been compared, any team member may initiate an identification poll. Only users who have completed all comparisons can vote, which helps ensure informed participation. Each user may cast one vote for their preferred match—or vote for “none” if no strong candidate emerges.

Poll results are displayed alongside each candidate (Figure~\ref{fig: poll_vote_section}). If at least half the members have voted and there is a clear majority --- suggesting a consensus was reached --- the project creator is given the option to finalize the match and publish the identity. If no consensus is reached, SleuthTalk displays alerts indicating a tie or inconclusive outcome, prompting further discussion.
\section{Evaluation} \label{se:Evaluation}

We conducted an exploratory, mixed-methods study to examine how SleuthTalk supports collaborative identification of unknown Civil War photographs in a private, structured workspace.

\subsection{Recruitment and Participants} \label{se:Recruitment and Participants}

We recruited six participants via CWPS and Civil War photography Facebook groups by sharing a demo video of SleuthTalk. Participants had varying levels of experience identifying Civil War photos (mean = 16 years, range = 0--44). Five participants identified as male and one as female, spanning age ranges from 18--20 to 51--60. Each received a \$50 Amazon gift card. We refer to participants as P1--P6.

\subsection{Procedure} \label{se:Procedure}

Each participant completed two identification tasks: one using SleuthTalk and one using a private Facebook group with over 13,000 members.\footnote{Group name anonymized to protect member privacy.} Tasks were assigned in alternating order over two weeks: P1--P3 began with SleuthTalk; P4--P6 began with Facebook. Tasks were swapped in the second week.

For the SleuthTalk task, participants uploaded one unknown Civil War photo to CWPS, created a project, shortlisted candidates, and optionally invited collaborators (CWPS users or non-users). Each project ran for one week. Participants were encouraged to use SleuthTalk’s full workflow, including comparison tools and the final poll.

For the Facebook task, participants posted a different unknown photo to the group and sought feedback. Because the group is private, we did not observe these interactions directly. Instead, participants described their experience during interviews.

\subsection{Data Collection and Analysis} \label{se:Data Analysis}

We analyzed system logs to examine user behavior on SleuthTalk, including project composition, candidate selection, voting, discussion, and poll activity. We also conducted semi-structured interviews with all six participants via Zoom. Interviews were transcribed and qualitatively analyzed around four themes: shortlisting, collaboration, structured feedback, and privacy.
\section{Findings}

In this section, we present findings from our mixed-methods evaluation of SleuthTalk. Drawing from participant interviews, usage logs, and user-generated annotations, we surface key themes that illustrate how SleuthTalk supported collaborative identification, influenced user behavior, and integrated into existing research workflows.

\subsection{Summary of Participant Projects}

To understand how participants used SleuthTalk during the study, we examined the nine projects created across the sessions. Each project varied in terms of team composition, number of candidates in the shortlist, amount of activity (e.g., votes, polls, and discussions), and whether a decision was reached. Most projects included invited collaborators, but acceptance rates varied. While some projects involved only the creator and one other member, others included nearly all study participants. In several cases, candidates received positive votes but still failed to reach consensus in the final poll. Only one project (Project 4) met the system threshold to finalize a photo ID, but the project creator did not mark the identity as confirmed. At the end of the study, all projects remained open. A summary of each project’s attributes is shown in Table~\ref{tab:project_summary}.

\begin{table*}[ht]
\centering
\resizebox{\textwidth}{!}{%
\begin{tabular}{|c|c|c|c|c|c|c|}
\hline
\textbf{Project} & \textbf{Creator} & \textbf{Members} (Joined/Invited) & \textbf{Candidates} & \textbf{Positive Votes} & \textbf{Poll Started} & \textbf{Poll Outcome} \\
\hline
1 & P2 & 2 / 3 & 2 & No & Yes & All "None" \\
2 & P2 & 1 / 1 & 3 & No & No & — \\
3 & P4 & 7 / 7 & 5 & Yes (4/29) & Yes & No consensus \\
4 & P4 & 6 / 6 & 3 & No & Yes & Meets threshold \\
5 & P3 & 6 / 6 & 3+ & Yes (later reversed) & Yes & Majority "None" \\
6 & P5 & 3 / 3 & 3 & No & No & — \\
7 & P1 & 7 / 9 & 3 & No & Yes & All "None" \\
8 & P6 & 3 / 5 & — & No & Yes & Tie \\
9 & P6 & 3 / 9 & 5 & Yes (2) & No & — \\
\hline
\end{tabular}%
}
\caption{Summary of SleuthTalk projects created by participants during the study. The table includes the project creator, number of members who joined and were invited, number of candidates in the shortlist, whether any positive votes were cast, whether a poll was started, and the poll outcome (if applicable).}
\label{tab:project_summary}
\end{table*}

\subsection{Shortlisting Potential Candidates} \label{se:Shortlisting Potential Candidates Finding}

\subsubsection{\textbf{Participants did not restrict themselves to the top AI matches when creating a shortlist}} \label{se:Participants did not restrict themselves to the top AI matches when creating a shortlist}

Usage logs show that participants selected candidates from a wide range of positions within the face recognition search results when building their shortlists. Out of the 37 candidates added across all nine projects, 8 were not among the top five matches returned by the AI system. Participants also chose candidates beyond the first few dozen results. The search interface in CWPS can return hundreds of potential matches, and several participants reported examining results across the full range. P4 explained, \textit{``I went through as many photos as I could; I did not just limit myself to the top.''}

During interviews, participants described using a combination of visual and contextual clues when deciding whether to add a candidate. These included facial resemblance, uniform details, name matches, and other biographical information. For example, P1 stated, \textit{``[I] also take into account what matches in [the] uniform.''} Other participants mentioned examining geographic region, military unit, or known affiliations based on the metadata displayed in the search results. In one case, a participant added a candidate with a facial similarity score below 0.50 by using the pre-identification search interface. This participant (P1) searched for a specific name and added a candidate whose middle name differed from the one originally searched. As P1 explained, \textit{``The name was part of the reason I added him because [I was] under the impression that [the last name of] the photo's [identity] is Wells.''}

\subsubsection{\textbf{Most participants did not add or archive candidates after the project was created}} \label{se:Most participants did not add or archive candidates to manage their shortlist}

Across all nine projects, only one participant added a new candidate after the project was created. According to the usage logs, this was P6, who explained during the interview, \textit{``I didn't have enough time when I first went through. I went back through it. Eventually, I saw one photo that might be [the one], and I added [him to the shortlist].''} Although no other participants added new candidates during the study, some described reasons they might do so in a longer-term setting. For instance, P3 mentioned discovering an alternate view of a candidate already on the shortlist and expressed interest in adding it: \textit{``So you could build up your argument a little bit more.''}

No participants used the archive feature during the study. This may be due to limited time or the feature's visibility in the interface. In interviews, all six participants stated that they would treat SleuthTalk projects as long-term efforts outside of the study setting, and would use features like adding or archiving candidates to manage evolving shortlists. One participant (P4) expressed interest in seeing votes on archived candidates to revisit earlier judgments. As P4 described, \textit{``If in the future, you invite somebody else, somebody new to the project, then they go through the ones that have already been there, and maybe they might see differently.''} Although SleuthTalk supports this functionality, it was not activated during the study, as no candidates were archived.

\subsection{Private Collaborative Workspace} \label{se:Private Collaborative Workspace Finding}

\subsubsection{\textbf{Participants invited a variety of collaborators to their workspace}} \label{se:Participants invited a variety of people to the workspace.}

Across the nine projects created during the study, participants invited a range of collaborators, including family members, friends, experts, and individuals they had not previously met. For example, P3 invited a family member, stating, \textit{``I also invited my wife, and she has her [CWPS] account as well."} Others invited friends to participate in their projects. P2, for instance, invited a personal friend to join both of his projects. Participants also invited domain experts. P6 explained that he included a professional from the Civil War photo community, stating, \textit{``He does this professionally, and you know so to have him in [the project] is very important."} In some cases, participants also invited individuals they had not previously interacted with. P6 mentioned inviting a stranger he had never met prior to the project.

\subsubsection{\textbf{Participants used polls to externalize tentative judgments and reconsider earlier assumptions}} \label{se:SleuthTalk helps with combating confirmation bias.}

Participants described using SleuthTalk’s poll mechanism to externalize their current thinking, even when they were uncertain. Several participants voted for a candidate in the poll despite expressing hesitation in the earlier identification vote. For instance, P5 noted, \textit{``I really became conflicted because there’s a lot that looks the same. After I considered it a little more, I thought I’m going to put my voice here that I think it’s him."} Similarly, P6 stated, \textit{``I thought that this [candidate] was the best match. I’m just not sure it’s a direct match."}

In Project 5, the poll became a site for participants to reflect on and revise their earlier judgments. Initially, P3 voted positively for a candidate who later received three negative votes and two ``not sure'' responses from other project members.  At the end of the study, P3 changed his mind and started a discussion thread, where he explained that``while I agree with <project member> that <candidate> seemed like the most promising, after thoroughly comparing all candidates to my image, I have to come to the conclusion that none of them is my officer." Later, in the interview, P3 told us that he went back to do more research about this candidate after receiving many counter-votes from other project members. He explained, ``\textit{I'd also found some other pictures of <candidate> when he was younger, without facial hair.}” P3 changed his mind after comparing with the new photo he found with the unknown photo.

\subsubsection{\textbf{Private projects supported open discussion while limiting public exposure of uncertain identifications}} \label{SleuthTalk's private projects provide a workspace for users to discuss the identity of the unknown photo, but also help to prevent the spread of misidentifications.}

Participants described being cautious about publicly sharing identifications without strong evidence. Several noted that they had prior experiences of misidentifying Civil War photos or observing misidentifications in public forums. P5 reflected, \textit{``[I try] not to identify someone if I am not 100\% sure because it's hard to undo that. And I actually did make that mistake once and it still haunts me."} Similarly, P2 stated, \textit{``I generally don't like to make something set in stone. This is who it is, unless there's a lot of evidence to support it. 'Cause I think it's kind of bad to misidentify people."}

Despite these concerns, participants often used the final poll to share their current preferences, even when they had earlier selected ``not sure'' in the identification vote. This behavior occurred in 4 projects (Projects 3, 4, 5, and 8) and involved 5 participants. For example, in Project 4, a candidate received four votes in the poll but had previously received six ``not sure'' votes during the identification voting process. Participants used the poll as a way to indicate a \emph{leaning}, not to confirm an identity. These intermediate preferences were not publicly visible during the study due to SleuthTalk’s private workspace model.

\subsubsection{\textbf{Participants valued peer input, but did not treat it as final}} \label{Other users' feedback is helpful but insufficient on its own to determine the final identity.}

Participants described SleuthTalk's peer feedback as useful for challenging assumptions and encouraging re-evaluation. Some noted that they had a tendency to focus on preferred candidates and that external perspectives helped counter this bias. P3 said, \textit{``Sometimes you get those blinders on because you want it to be that person so bad. \ldots If a couple of other people say something, it gives you a second look, and helps you."}

At the same time, participants reported treating others’ opinions as suggestions rather than conclusions. For example, in Project 4, even though four of six participants voted for the same candidate in the poll, the project creator chose not to finalize the identification. P5 explained, \textit{``I do take that into consideration, and I might be the only dissenting voice. I kind of want to still leave it open for interpretation. I think the more evidence you have, it very much helps the photo."}

\subsubsection{\textbf{Participants valued private workspaces for reducing social friction and protecting sensitive materials}} \label{se:The controlled private workspace helped to prevent toxic comments and counterfeiting.}

Participants shared prior experiences with public forums and social media platforms that included toxic interactions, distractions, and concerns over intellectual property. Several felt that SleuthTalk’s private, controlled workspace provided a more focused and respectful environment.

For instance, P2 described receiving offensive messages when posting in Facebook Civil War groups: \textit{``Oftentimes it is a very toxic community on Civil War forums, where they are like, [they] know everything. And then get into fights about it in the comments. I've had people send horrible messages to me, people get really heated about stuff on [Facebook]."} In contrast, he explained that using SleuthTalk with a small, known group helped reduce these issues: \textit{``I think that definitely would be eliminated in a forum like [SleuthTalk] because you can just have people that you know, aren't going to go like nuts."}

Participants also expressed concerns about losing control of valuable or rare photo artifacts in public settings. P6 explained, \textit{``Someone just sort of right-clicks and takes your image or does a screenshot and take your image and uses it to make a counterfeit copy or use it in their publication or use it without attribution."} P6 also noted safety concerns with sharing expensive items: \textit{``It's not always ideal to just stick [the photo] you have that's worth some money on a public forum and everyone now knows you have it."} Similarly, P4 described how collectors often avoid public posting to limit exposure of valuable private collections.

Participants noted that SleuthTalk’s invitation-based model offered more control over who accessed their projects. P6 emphasized the difference in participation quality between CWPS and broader platforms like Facebook: \textit{``You have to sort of show the commitment [to use the website]."} Participants perceived the CWPS community—and by extension, SleuthTalk—as a space with a higher level of focus and shared interest.

\subsection{Structured Feedback} \label{se:Structured Feedback Finding}

\subsubsection{\textbf{Participants used both comparison and identification votes}} \label{se:Users voted on both individual comparisons and across candidates.}

Usage logs showed that participants engaged extensively with the structured feedback tools in SleuthTalk. Across all nine projects, participants submitted a total of 152 facial feature comparison votes and 125 identification votes. In addition, 48\% of project members voted in at least one poll, selecting the candidate they believed was the best match from the shortlist.

\subsubsection{\textbf{Participants found SleuthTalk’s structured workflow easy to follow and informative}} \label{se:Users find that SleuthTalk provides an organized and structured way to see others' opinions.}

SleuthTalk guided participants through a multi-step process, starting with feature-level comparisons and identification voting, followed by discussion and polling. Logs confirmed that all participants completed the comparison and identification voting steps in their respective projects.

Participants described the structured workflow as intuitive and useful for understanding peer input. As P5 described, \textit{``It was easy to understand how it was, how it is layered and how it is structured\ldots There's mechanisms in place to measure and there's opportunities to get the quantitative information but also have the anecdotal discussion."}

In total, participants made 254 facial feature selections across all comparisons. Of these, 89\% corresponded to predefined categories including hair, nose, eye(s), eyebrow, jawline, facial hair, and ear(s). The distribution of these selections is shown in Table~\ref{table:facial_feature_selection}.

Not all projects proceeded to the polling stage. In some cases, participants did not start a poll when no viable candidates emerged. For example, P2 explained, \textit{``I didn't think it was really necessary. Because it seems everyone is pretty confident that they weren't [the match].''}

\begin{table*}[t]
\centering
\begin{tabular}{|c|c|c|c|c|c|c|c|}
\hline
\textbf{Feature} & Hair & Nose & Eye(s) & Eyebrow & Jawline & Facial Hair & Ear(s) \\
\hline
\textbf{Count} & 47 & 40 & 33 & 31 & 29 & 24 & 20 \\
\hline
\end{tabular}
\caption{Usage of predefined facial feature categories across all comparison votes}
\label{table:facial_feature_selection}
\end{table*}

\subsubsection{\textbf{Participants found SleuthTalk discussions more focused than on public forums}} \label{Users think SleuthTalk is more focused and helps to filter out irrelevant information.}

Several participants compared SleuthTalk to their previous experiences using Facebook groups for Civil War photo identification. Participants noted that public discussions on Facebook often lacked focus or relevance. P3 explained, \textit{``It just seems like sometimes people miss the mark, they just kind of trail off after a while."} Similarly, P4 noted that, \textit{``You might have some totally random person that just happens to be a forum member\ldots they might leave a [reply] in their comment, and [it] might not even be relevant to the actual discussion."}

Beyond off-topic commentary, some participants reported frustration with commercial messages. P5 shared, \textit{``It's just frustrating to me [that] people will message you [if they] want to buy things from you. Sometimes it's not a great experience."}

In contrast, participants reported that SleuthTalk’s private, structured setting led to more relevant and constructive discussion. P6 attributed this to the ability to invite a trusted group: \textit{``The discussion is helpful [in SleuthTalk] because I think you're sort of ensuring a higher quality of people that are in the discussion."} P5 added, \textit{``One thing I really enjoyed about it was [that] it sort of default filters out the irrelevant conversation."}

\subsection{Perceptions of SleuthTalk vs. Facebook} \label{se:SleuthTalk vs. Facebook}

\subsubsection{\textbf{Participants viewed SleuthTalk and Facebook as serving different but complementary purposes}} \label{Participants think that SleuthTalk serves different, but complementary, purposes from Facebook groups.}

During interviews, several participants compared using SleuthTalk and Facebook groups during the study. Half of the participants mentioned that they received potential identifications for their unknown photo through the Facebook task. They attributed this success to the large membership of the Civil War Facebook group, which has over 13,000 members.

Participants found Facebook particularly helpful for generating initial leads when they had no starting information. P2 explained, \textit{``I just want to get a broad sense if anyone recognizes, like this guy, or has a photo of this guy in their collection."} Similarly, P4 described how specific collectors within the Facebook group are known for particular regions or photo types: \textit{``<collector's name on Facebook> knows New Hampshire images, so if you have an image that has a backmark from New Hampshire, you would have a good chance running it by him, because it's very possible he might know who that guy is, just by memory, because he has been studying."}

In contrast, participants saw SleuthTalk as more suitable for comparing known candidates in a focused setting. Several participants noted that once they had narrowed their search to a few plausible matches, SleuthTalk was their preferred platform for discussion and peer input. P2 described this use case: \textit{``If I have three [candidates] that I think are all really like [the photo I uploaded], I'm gonna ask very direct people get their opinion on just the faces. And I don't need historical context or to reach out to wider collectors or anything, [such as the Facebook group]."}

Some participants reported using the two platforms in tandem—first crowdsourcing candidate suggestions on Facebook, then moving to SleuthTalk to facilitate more structured and collaborative deliberation.

\subsubsection{\textbf{Participants noted that Facebook may produce faster responses, but more inconsistent engagement}} \label{Participants think it is faster to receive a response from the Facebook group than SleuthTalk, but the request is more likely to be ignored on Facebook.}

Several participants reported that Facebook often yielded faster responses than SleuthTalk. For example, P4 said, \textit{``I was surprised because it got identified really quickly."} P1 attributed the speed to Facebook’s broader visibility and user base: \textit{``Facebook is more well-known than [the SleuthTalk] at this time, so if they're on Facebook, that's going to be an automatic, fast response. [But for SleuthTalk], it might be a slower response."}

P1 also pointed out that SleuthTalk participation is more limited by design, since it requires users to join private projects, and that responses might take longer if people were not actively checking for updates: \textit{``Unless you're in the project or used to the [website], just because people might be working and they'll have to see the email, to [join the project and] respond."}

At the same time, P1 noted that even though Facebook has more users, posts could be ignored, especially if the user was less familiar to the community: \textit{``People responded on the Facebook page, and it seems like a lot of them. [But] I put a comment out there, I didn't always get a response, maybe the other members knew each other or had interacted before."} In contrast, P1 felt that she was more likely to receive a response in SleuthTalk because she had more control over who was invited: \textit{``I was kind of going for a targeted audience of people that I knew had an interest in this kind of thing."}
\section{Discussion}

\subsection{From Open Crowdsourcing to Trusted Micro-Crowds}

SleuthTalk redefines the traditional approach to historical photo identification by transitioning from large, open crowdsourcing platforms to smaller, trust-based collaborations. While platforms like Facebook offer broad reach, our participants highlighted challenges such as disorganized discussions~\cite{mohanty2019second}, off-topic comments~\cite{zhang2017wikum}, and concerns about reputational risks associated with public misidentifications. SleuthTalk addresses these issues through its invitation-based model, allowing users to curate focused groups of trusted collaborators, including friends, family, and domain experts. This 
"micro-crowd" approach fosters a more controlled and supportive environment for deliberation. Our findings indicate that participants felt safer and more focused within SleuthTalk’s private setting, noting a reduction in irrelevant conversations and negative interactions. Prior work suggests that in many real-world tasks, moderately sized groups may strike a balance between diversity and coherence, supporting effective collective decisions~\cite{galesic2018smaller}. SleuthTalk builds on this intuition by supporting structured collaboration among trusted micro-groups. By introducing an intermediate, private deliberation step into the CWPS workflow, SleuthTalk allows users to engage in thoughtful analysis before making identifications public, with the goal of improving the overall accuracy and reliability of the identification process.

\subsection{Trust, Uncertainty, and the Role of Privacy}

SleuthTalk’s private, invite-only model not only shapes who participates in collaborative identification, but also changes how uncertainty is handled. Participants used SleuthTalk’s workspace to express tentative opinions, revisit earlier judgments, and voice disagreement --- actions that might carry reputational or social risk in public forums. Our findings show that users often withheld strong conclusions until they had reviewed others’ reasoning and conducted structured comparisons. In several cases, participants revised their votes or retracted earlier identifications after discussion. One participant explained that external disagreement prompted them to re-examine evidence and ultimately vote “none of the above.” This kind of epistemic humility was supported by SleuthTalk’s poll mechanism and discussion features, which allowed users to externalize incomplete thinking without prematurely finalizing a claim. By embedding deliberation before public exposure, SleuthTalk reframes identification as a process of collective reflection --- not just assertion --- allowing users to navigate uncertainty with greater care and confidence.

At the same time, private micro-crowds can introduce important tradeoffs. While limiting public exposure may reduce reputational pressure and make it easier for participants to express uncertainty, small-group deliberation can also be shaped by social dynamics that produce homogenization or reinforce shared assumptions. Our exploratory study did not evaluate the potential risks that may be compounded when deliberation begins from algorithmically generated candidate matches. Prior evaluations have documented demographic performance disparities in face recognition systems~\cite{buolamwini2018gender}. Further, research on AI-assisted group decision-making suggests that groups may over-rely on algorithmic recommendations unless mechanisms explicitly encourage critical reflection~\cite{chiang2024enhancing}. Future work should examine how group composition, algorithmic recommendations, and mechanisms for selectively sharing deliberative outcomes influence bias, accountability, and decision quality.

\subsection{Structured Sensemaking as Collective Intelligence}

SleuthTalk supports structured sensemaking by guiding users through a workflow that surfaces reasoning, organizes evidence, and facilitates deliberation. Participants engaged deeply with the comparison tools, submitting over 250 facial feature annotations and 125 identity votes across nine projects. These annotations made individual judgments explicit: cropped regions of jawlines, eyes, or ears revealed what users were noticing, and prompted others to re-evaluate their own assumptions. Threaded discussions added a space for open-ended reasoning, allowing participants to explain, challenge, or revise interpretations. In one case, a participant used the thread to publicly retract an earlier vote after comparing perspectives. Polls and visual summaries enabled groups to track emerging consensus without forcing premature closure. Participants often chose “none” or “not sure,” treating uncertainty as an acceptable outcome. This modular workflow, separating fine-grained visual comparison, individual judgment, and group deliberation, offers a reusable design pattern for complex collaborative decision-making tasks. Together, these tools often structured the identification process as an iterative, social inquiry, and in several cases, scaffolding attention, transparency, and dialog helped facilitate more rigorous, collective decision-making. 

\subsection{Design Implications Beyond Historical Photos}

SleuthTalk’s design principles can be transferable and have broader implications beyond historical photo identification. By providing a private, structured space for collaboration, SleuthTalk addresses common challenges in many domains involving complex or ambiguous data. For example, commercial genealogy platforms like Ancestry.com have recently begun exploring private spaces for users to engage in collaborative research~\cite{AncestryNetworks}, allowing trusted collaborators to share and discuss historical records without the distractions or reputational risks of open forums. Yet, little is known about how effective these private research environments are or what challenges users face when moving from open crowdsourcing to more controlled, smaller-scale collaboration. SleuthTalk offers valuable insights into how these platforms can be structured to facilitate deeper, more reflective collaboration.

Similarly, domains like legal decision-making, medical diagnostics, or digital forensics could benefit from SleuthTalk’s model, where participants weigh evidence iteratively, collaborate without fear of judgment, and refine conclusions through structured deliberation. SleuthTalk projects also serve as work artifacts capturing comparisons, discussion threads, and poll outcomes that can be revisited or --- with additional features for exporting and/or monitoring --- audited and used as teaching tools. Looking ahead, there is also potential to integrate large language models (LLMs) into these workflows to assist with summarizing discussion, highlighting disagreement, or prompting new angles of analysis~\cite{10.1145/3613904.3642530}. By grounding decision-making in transparency, shared context, and thoughtful engagement, SleuthTalk’s design could be a stepping stone for broader applications that require collective intelligence and careful, accountable decision-making.
\section{Conclusion and Future Work}

We introduced SleuthTalk, a private, structured workspace designed to support collaborative identification of historical photographs. The system addresses key limitations of open platforms by enabling trusted micro-crowds to deliberate, compare, and reason through ambiguous matches without public pressure. Our mixed-methods evaluation with six experienced photo researchers suggests that SleuthTalk supports self-reported confidence, encourages epistemic humility, and generates reviewable artifacts of group reasoning. This approach opens doors to new models of collaborative sensemaking in domains like genealogy, forensics, and medical diagnostics. While our study was limited in scale, future work could deploy SleuthTalk more broadly on CWPS to examine sustained, in-the-wild collaboration and understand engagement at community scale.

\begin{acks}
    We would like to thank our participants and the reviewers for their helpful feedback. This research was supported in part by NSF IIS-1651969 and a Virginia Tech ICTAS Junior Faculty Award.
\end{acks}

\bibliographystyle{ACM-Reference-Format}
\bibliography{sample-base}




\end{document}